\documentclass[aps,prd,preprint,superscriptaddress,showpacs,floatfix,nofootinbib]{revtex4}

\usepackage{graphicx}

\newcommand{\bmat}{\left(\begin{array}}
\newcommand{\emat}{\end{array}\right)}
\newcommand{\beq}{\begin{equation}}
\newcommand{\eeq}{\end{equation}}

\def\yzero{\smash{\hbox{$y\kern-4pt\raise1pt\hbox{${}^\circ$}$}}}

\def\-{\hphantom{-}}

\def\s2{\frac{1}{\sqrt2}}

\def\be{\begin{equation}}
\def\ee{\end{equation}}
\def\bea{\begin{eqnarray}}
\def\eea{\end{eqnarray}}

\def\IF{\relax{\rm I\kern-.18em F}}
\def\II{\relax{\rm I\kern-.18em I}}
\def\IP{\relax{\rm I\kern-.18em P}}

\def\Dsl{\,\raise.15ex\hbox{/}\mkern-13.5mu D} 

\def\IC{\bf C}
\def\IZ{\bf Z}

\def\z2z2{$\IC^3/(\IZ_2\times\IZ_2)$}

\def\s{\sigma}

\def\z{\zeta}

\def\bo{{\raise-.3ex\hbox{\large$\Box$}}}               
\def\face{{\raise.2ex\hbox{$\displaystyle \bigodot$}\mskip-2.2mu \llap {$\ddot
        \smile$}}}                                      

\def\leftrightarrowfill{$\mathsurround=0pt \mathord\leftarrow \mkern-6mu
        \cleaders\hbox{$\mkern-2mu \mathord- \mkern-2mu$}\hfill
        \mkern-6mu \mathord\rightarrow$}       
\def\dvec#1{\vbox{\ialign{##\crcr
        \leftrightarrowfill\crcr\noalign{\kern-1pt\nointerlineskip}
        $\hfil\displaystyle{#1}\hfil$\crcr}}}           

\def\beq{\begin{equation}}
\def\eeq{\end{equation}}

\def\beqx{\begin{displaymath}}
\def\eeqx{\end{displaymath}}

\def\beqa{\begin{eqnarray}}
\def\eeqa{\end{eqnarray}}

\begin{document}

\title{ Multiple D2-Brane Action from M2-Branes }

\author{Tianjun Li}

\affiliation{Institute of Theoretical Physics, 
Chinese Academy of Sciences, Beijing 100080, P. R. China }

\affiliation{George P. and Cynthia W. Mitchell Institute for
Fundamental Physics, Texas A$\&$M University, College Station, TX
77843, USA }

\author{Yan Liu}

\affiliation{Institute of Theoretical Physics, 
Chinese Academy of Sciences, Beijing 100080, P. R. China }

\author{Dan Xie}

\affiliation{George P. and Cynthia W. Mitchell Institute for
Fundamental Physics, Texas A$\&$M University, College Station, TX
77843, USA }

\date{\today}

\begin{abstract}

We study the detail derivation of the multiple D2-brane effective 
action from multiple M2-branes in the Bagger-Lambert-Gustavsson 
(BLG) theory and the Aharony-Bergman-Jafferis-Maldacena (ABJM) theory
by employing the novel Higgs mechanism. We show explicitly
that the high-order $F^3$ and $F^4$ terms are commutator terms, 
and conjecture that all the high-order terms are commutator terms. 
Because the commutator terms can be treated as the covariant 
derivative terms, these high-order terms do not contribute to the 
multiple D2-brane effective action. Inspired by the derivation 
of a single D2-brane from a M2-brane, we consider the curved M2-branes 
and introduce an auxiliary field. Integrating out the auxiliary field, 
we indeed obtain the correct high-order $F^4$ terms in the D2-brane 
effective action from the BLG theory
and the ABJM theory with $SU(2)\times SU(2)$ gauge symmetry,
but we can not obtain the correct high-order $F^4$ terms 
from the ABJM theory with $U(N)\times U(N)$ and
$SU(N)\times SU(N)$ gauge symmetries for $N > 2$. We also briefly
comment on the (gauged) BF membrane theory.

\end{abstract}


\preprint{MIFP-08-16}

\maketitle



\section{Introduction}

Inspired by the ideas that the Chern-Simons gauge theories without
Yang-Mills kinetic terms may be used to describe  
${\cal N}=8$ superconformal M2-brane world-volume 
theory~\cite{0411077, 0412310},
Barger and Lambert~\cite{0611108, 0711.0955, 0712.3738}, as well as 
Gustavasson (BLG)~\cite{0709.1260,0802.3456} have successfully 
constructed three-dimensional ${\mathcal N}=8$ superconformal 
Chern-Simons gauge theory 
with manifest $SO(8)$ R-symmetry based on three algebra.  And then
there is intensive research on the world-volume action of
 multiple coincident M2-branes~\cite{Mukhi:2008ux, 0803.3242, 0803.3611,
0803.3803, 0804.0913, 0804.1114, 0804.1256, 0804.1784, 0804.2110,
0804.2186, 0804.2201, 0804.2519, 0804.2662, 0804.3078, 0804.3567,
0804.3629, 0805.1012, 0805.1087, 0805.1202, 0805.1703, 0805.1895,
0805.1997, 0805.2898, 0805.3125, 0805.3193, 0805.3236, 0805.3427,
0805.3662, 0805.3930, 0805.4003, 0805.4363, 0805.4443, 0806.0054,
0806.0335, 0806.0363, 0806.0738, 0806.1218, 0806.1420, 0806.1519,
0806.1639, 0806.1990, 0806.2270, 0806.2584, 0806.3242, 0806.3253,
0806.3391, 0806.3498, 0806.3520, 0806.3727, Minahan:2008hf,
Larsson:2008ke, Furuuchi:2008ki, Armoni:2008kr, Agarwal:2008rr,
Gaiotto:2008cg, Bandos:2008fr, 0806.4807, Bedford:2008hn,
Arutyunov:2008if, Stefanski:2008ik, Grignani:2008is, Hosomichi:2008jb,
Fre:2008qc, Okuyama:2008qd, Bagger:2008se, Terashima:2008sy,  
Grignani:2008te, Chakrabortty:2008zk, Terashima:2008ba, Gromov:2008bz,
Ahn:2008gd, 0807.0777, 0807.0802, 0807.0808, 0807.0812, 0807.0880,
0807.0890, 0807.1074}.  Although the 
BLG theory is expected to describe any number of M2-branes, 
there is one and only one known example with gauge group $SO(4)$ for 
the positive definite metric~\cite{0804.2110, 0804.2662, 0804.3078}. 
At the level one of the Chern-Simons gauge theory, the BLG 
SO(4) gauge theory describes two M2-branes on a $R^8/Z_2$ 
orbifold~\cite{0804.1114, 0804.1256}.
Thus, it is very important to generalize the BLG theory so that
it can describe an arbitrary number of M2-branes.

By relaxing the requirement of the positive definite metric on three
algebra, three groups~\cite{0805.1012,0805.1087,0805.1202}
proposed the so called BF membrane theory with 
arbitrary semi-simple Lie groups. However, the BF membrane theory has 
ghost fields and then the unitarity problem in the classical theory 
due to the Lorenzian three algebra. To solve these problems,
 the global shift symmetries for the bosonic and fermionic ghost 
fields with wrong-sign kinetic terms are gauged, which ensures 
the absence of the negative norm states in the 
physical Hilbert space~\cite{0806.0054,0806.0738}.
However,  this gauged BF membrane theory might be equivalent to 
three-dimensional ${\cal N}=8$ supersymmetric Yang-Mills 
theory~\cite{0806.1639} via a duality transformation 
due to de Wit, Nicolai and Samtleben~\cite{deWit:2004yr}.

Very recently, Aharony, Bergman, Jafferis and Maldacena (ABJM)
have constructed three-dimensional Chern-Simons theories
with gauge groups $U(N)\times U(N)$ and $SU(N)\times SU(N)$ 
which have explicit ${\cal N}=6$ superconformal 
symmetry~\cite{0806.1218} (For Chern-Simons gauge theories with 
${\cal N}=3$ and $4$ supersymmetries, see 
Refs.~\cite{0704.3740, Gaiotto:2008sd}). Using
brane constructions they argued that the  $U(N)\times U(N)$
theory at Chern-Simons level $k$ describes the low-energy 
limit of $N$ M2-branes on a $C^4/Z_k$ orbifold. In particular,
 for $k=1$ and $2$, ABJM conjectured that their theory
describes  the $N$ M2-branes respectively in the flat space and 
on a $R^8/Z_2$ orbifold, and then might have ${\cal N}=8$ supersymmetry.
For $N=2$, this theory has extra symmetries and is the same as
 the BLG theory~\cite{0806.1218}.

On the other hand, D-branes are the hypersurfaces on which the 
open strings can end, and their dynamics is described by open string
field theory~\cite{Polchinski:1995mt}. The low-energy world-volume 
action for D-branes can 
be obtained by calculating the string scattering
amplitudes~\cite{Tseytlin:1997csa} or by using the 
T-duality~\cite{Myers:1999ps}. As usual in string theory,
there are high-order $\alpha'=\ell_s^2$ corrections, where
$\ell_s$ is the string length scale. For a single D-brane,
the D-brane action, which includes all order corrections in
the gauge field strength but not its derivatives, takes the 
Dirac-Born-Infeld (DBI) form~\cite{Leigh:1989jq}. 
For multiple coincident D-branes, Tseytlin assumed that
all the commutator terms should be treated as covariant 
derivative terms for gauge field strength, and thus should not be
included in the effective action~\cite{Tseytlin:1997csa}. 
And he proposed that 
the action is the symmetrized trace of the direct non-Abelian 
generalization of the DBI action~\cite{Tseytlin:1997csa}. 
This non-Abelian DBI action gives the correct terms up to 
the order $F^4$ that were completely determined 
previously~\cite{Tseytlin:1986ti, Gross:1986iv}. But it 
fails for the higher order 
terms~\cite{Hashimoto:1997gm, Schwarz:2001ps}. Because 
the $F^3$ terms can always be written 
as the commutator terms, they are not interesting in
the discussions of the D-brane effective action.

With the multiple M2-brane and D2-brane theories, we can
study the deep relation between them. As we know, the full 
effective action of a D2-brane can be obtained by the reduction of 
the eleven-dimensional supermembrane action~\cite{Bergshoeff:1996tu}. 
So, whether we can obtain the effective non-Abelian action 
for multiple D2-branes from the reduction of the BLG  and ABJM 
theories is an interesting open question. Mukhi and Papageorgakis 
proposed a novel Higgs mechanism by giving vacuum expectation
value (VEV) to a scalar field, which can promote the topological 
Chern-Simons gauge fields to dynamical gauge 
fields~\cite{Mukhi:2008ux}. And they indeed obtained the 
maximally supersymmetric Yang-Mills theory for two D2-branes 
from the BLG theory at the leading order. Also, there exists 
a series of high-order corrections~\cite{Mukhi:2008ux}.

In this paper, we consider the derivation of the multiple 
D2-brane effective action from the multiple M2-branes in the 
BLG and ABJM theories in details. Concentrating on pure 
Yang-Mills fields,
we show that the high-order $F^3$ and $F^4$ terms are 
 commutator terms, and argue that all the high-order terms 
are also commutator terms. Thus, these high-order 
terms are irrelevant to the multiple D2-brane effective action.
Note that the (gauged) BF membrane theory does not have high-order
terms, the BLG theory, the (gauged) BF membrane theory, and 
the ABJM theory give the same D2-brane  effective action.
In order to generate the non-trivial high-order $F^4$ terms, 
inspired by the derivation of a single D2-brane 
from a M2-brane~\cite{Bergshoeff:1996tu},
we consider the curved M2-branes and introduce an auxiliary
field. In particular, the VEV of the scalar field in the novel 
Higgs mechanism depends on the auxiliary field. After we
integrate out the massive gauge fields and auxiliary field, 
we indeed obtain the high-order 
$F^4$ terms in the D2-brane effective action from the BLG  theory
and the ABJM theory with $SU(2)\times SU(2)$ gauge group.
However, we still can not
obtain the correct $F^4$ terms in the generic ABJM theories
with gauge groups $U(N)\times U(N)$ and $SU(N)\times SU(N)$ 
for $N>2$. The reason might be that
  the $SU(2)\times SU(2)$ gauge theory has 
three-dimensional ${\cal N}=8$ superconformal symmetry while the 
$U(N)\times U(N)$ and
$SU(N)\times SU(N)$ gauge theories with $N > 2$ may only have 
three-dimensional ${\cal N}=6$ superconformal 
symmetry~\cite{Bagger:2008se}.
We also briefly comment on the  (gauged) BF membrane theory.

This paper is organized as follows. In Section II, we
briefly review the novel Higgs mechanism in the BLG theory and 
(gauged) BF  membrane theory,  and study the novel Higgs mechanism in
the ABJM theory.  In Section III, we calculate the effective D2-brane
action with the leading order $F^2$, and 
high-order $F^3$ and $F^4$ terms from M2-branes. In Section IV, 
we generate the high-order
 $F^4$ terms by considering the curved M2-branes and introducing an 
auxiliary field. Our discussion and conclusions are given in Section V.

\section{Novel Higgs Mechanism }

In this Section, we briefly review the novel Higgs mechanism from
 M2-branes to D2-branes in the BLG theory and (gauged) BF membrane 
theory, and study it in the ABJM theory. 

\subsection{The BLG Theory and BF Membrane Theory}

In the  Lagrangian for the BLG theory with gauge group 
$SO(4)$~\cite{0711.0955}, we define
\begin{eqnarray}
f^{abcd} ~\equiv~ f \epsilon^{abcd}~,~~~  
f=\frac{2\pi}{k}~,~\,
\end{eqnarray}
where $k$ is the level of the Chern-Simons terms.
We also make the following transformation on the
Yang-Mills fields
\begin{eqnarray}
A_{\mu AB} \longrightarrow {1\over f} A_{\mu AB}~.~\,
\end{eqnarray}
Then the Lagrangian for the BLG theory with gauge group $SO(4)$ 
becomes
\begin{eqnarray}
{\mathcal L}&=&-{1\over2}D^{\mu}X^{AI}D_{\mu}X^I_A+{i\over2}\bar{\Psi}^A\Gamma^\mu
D_\mu\Psi_A+{{if}\over4}\bar{\Psi}_B\Gamma_{IJ}X_C^IX_D^J\Psi_A\epsilon^{ABCD} \nonumber\\
&& -V(X)+{1\over {2f}}\epsilon^{\mu\nu\lambda}(\epsilon^{ABCD}A_{\mu
AB}\partial_{\nu}A_{\lambda
CD}+{2\over3}{\epsilon^{CDA}}_G\epsilon^{EFGB}A_{\mu AB}A_{\nu
CD}A_{\lambda EF})~,~
\label{LAG-BLG}
\end{eqnarray}
where $A=1,2,3,4$, $I=1,2,...,8$, and 
\begin{equation}
V(X)={{f^2}\over12}\epsilon_{ABCD}
{\epsilon_{EFG}}^DX^{A(I)}X^{B(J)}X^{C(K)}X^{E(I)}X^{F(J)}X^{G(K)}~.~\,
\end{equation}
As we know, the strong coupling limit of Type IIA theory is M-theory,
and the coupling constant in Type IIA theory is related to the radius of 
the circle of the eleventh dimension in M-theory. Thus, for D2-branes, 
the gauge coupling constant is also
related to the radius of the circle of the eleventh dimension. And at the strong
coupling limit the D2-branes become M2-branes. To derive the D2-branes from
M2-branes via the novel Higgs mechanism, we compactify the M-theory on the circle 
of the eleventh dimension 
by giving VEV to a linear combination of the scalar fields 
$X^{A(I)}$~\cite{Mukhi:2008ux}.
Because we have the $SO(8)$ R-symmetry and SO(4)
gauge symmetry, we can always make the rotation so that
only the component $\langle X^{8(\phi)} \rangle $ develops a VEV 
\begin{equation}
 \langle X^{8(\phi)} \rangle ~=~v_0~=~{v \over {\sqrt f}} ~,~\,
 \end{equation}
where we split the index $A$ into two sets $a=1,2,3$ and $\phi=4$.
In addition, the gauge fields are
splitted into $A_{\mu}^a$ and $B_{\mu}^a$
\begin{equation}
A_\mu^a \equiv  A_{\mu}^{a\phi} ~,~~~
B_\mu^a \equiv {1\over2}{\epsilon^a}_{bc}A_\mu^{bc} ~.~\,
\end{equation}
And then the Chern-Simons terms can be rewritten as  
\begin{equation}
{1\over2}\epsilon^{\mu\nu\lambda}\epsilon^{ABCD}A_{\mu
AB}\partial_{\nu}A_{\lambda
CD}=4\epsilon^{\mu\nu\lambda}B_{\mu}^a\partial_{\nu}A_{\lambda a}~,~\,
\end{equation}
\begin{equation}
{1\over3}\epsilon^{\mu\nu\lambda}
{\epsilon_{CDA}}^G\epsilon_{EFGB}A_{\mu}^{ AB}A_{\nu}^{
CD}A_{\lambda}^{
EF}=-4\epsilon^{\mu\nu\lambda}\epsilon_{abc}B_{\mu}^aA_\nu^bA_\lambda^c-{4\over3}\epsilon^{\mu\nu\lambda}
\epsilon_{abc}B_\mu^aB_\nu^bB_\lambda^c~,~\,
\end{equation}
where we neglect the total derivative term.
Combining these two terms, the Chern-Simons action becomes
\begin{equation}
{\mathcal L}_{\rm CS}= {1 \over f} \left(
2\epsilon^{\mu\nu\lambda}B_\mu^aF_{\nu\lambda
a}-{4\over3}\epsilon^{\mu\nu\lambda}\epsilon_{abc}B_{\mu}^aB_{\nu}^bB_\lambda^c \right)~,~\,
\end{equation}
where $F_{\nu\lambda a}=\partial_{\nu}A_{\lambda a}-\partial_{\lambda}A_{\nu a}
-2\epsilon_{abc}A_\nu^bA_\lambda^c$ is
the field strength for the gauge field $A_\mu^a$. Similarly, the
kinetic terms for the scalar fields are
\begin{equation}
D_\mu X^{a(I)}=\partial_\mu
X^{a(I)}+\epsilon^a_{~BCD}A_\mu^{CD}X^{B(I)}=\partial_\mu
X^{a(I)}-2{\epsilon^a}_{cb}A_\mu^cX^{b(I)}+2B_\mu^aX^{\phi(I)},
\end{equation}
\begin{equation}
D_\mu X^{\phi(I)}=\partial_\mu X^{\phi(I)}-2B_{\mu a}X^{a(I)}.
\end{equation}
Substituting these back into the action and setting $X^{\phi(8)}\rightarrow
X^{\phi (8)}+v$, we obtain the terms involving $B_\mu^a$ from the scalar 
kinetic terms 
\begin{eqnarray}
{\mathcal
L}=-{{2v^2}\over f} B_\mu^aB_a^\mu
-{{4v}\over {\sqrt f}} X^{8\phi}B_\mu^aB_a^\mu-2X^{8\phi}X^{8\phi}B_\mu^aB_a^\mu
-2B_\mu^aB_a^\mu
X^{\phi(i)}X_{\phi(i)}\nonumber\\-2B_\mu^aX^{\phi(i)}D^{\mu}X_a^{(i)}
-{{2v}\over {\sqrt f}}  B_\mu^aD^\mu
X_a^{(8)}-2X^{8\phi}B_\mu^aD^\mu
X_a^{(8)}\nonumber\\
-2B_{\mu a}X^{a(I)}B_b^\mu X^{b(I)}+2B_a^\mu X^{a(I)}\partial_\mu
X^{\phi(I)},
\end{eqnarray}
where $i=1,2,...,7$, and the new defined
covariant derivative is $D_\mu
X^{a(I)}=\partial_{\mu}X^{a(I)}-2{\epsilon^a}_{bc}A_\mu^bX^{c(I)}$.
Therefore, the relevant Lagrangian for pure Yang-Mills fields is 
\begin{eqnarray}
\label{a} 
{\mathcal L}_{\rm YM}={1\over f} \left(
-2v^2B_\mu^aB_a^\mu+2\epsilon^{\mu\nu\lambda}B_\mu^aF_{\nu\lambda
a}-{4\over3}\epsilon^{\mu\nu\lambda}\epsilon_{abc}B_\mu^aB_\nu^bB_\lambda^c \right)~.~\,
\end{eqnarray}

Next, we would like to briefly review the result of the novel Higgs mechanism 
in the BF membrane theory~\cite {0805.1012,0805.1087,0805.1202}. Here, we follow 
the convention in Ref.~\cite{0805.1087} except that we choose
\begin{eqnarray}
(B_{\mu})_{a}~\equiv~ {1\over 2} (A_{\mu})_{b c} f^{~~b c}_{a}~.~\,
\end{eqnarray}
 In this theory, the equation of motion for ghost field $X_{-}^I$ gives the constraint 
$\partial^2 X^I_+=0$. So, we can give a constant VEV to $X^{8}_+$, {\it i.e.}, 
$X^{8}_+=v$. And then we obtain the relevant Lagrangian for pure gauge fields
\be
\label{b} 
{\mathcal L}=-2v^2B^a_\mu B^\mu_a+2\epsilon^{\mu\nu\lambda} B^a_\mu
F^a_{\nu\lambda} ~.~\, 
\ee 
It should be noted that unlike the Lagrangian in Eq. (\ref a) in the BLG theory,
there is no cubic term for $B_{\mu}^a$ in above Lagrangian.  And this 
is one of the motivations of the work~\cite{0806.1639} which showed that the
gauged BF membrane theory might be equivalent to the maximally 
supersymmetric three-dimensional Yang-Mills theory via
 a duality transformation due to de Wit, Nicolai and 
Samtleben~\cite{deWit:2004yr}.

After gauging the shift symmetries for the ghost fields
$X^I_-$ and $\Psi_-$ in the BF membrane
theory~\cite{0806.0054,0806.0738} by introducing new gauge fields, 
we could make the gauge choice to decouple the ghost states.
And the equation of motion for the new gauge fields gives 
the constraint $\partial_\mu X^I_+=0$,
which indicates that $X^I_+$ must be a constant. 
We emphasize that in this case the relevant Lagrangian for 
 pure Yang-Mills fields is still given by Eq. (\ref b).
 
\subsection{ The ABJM Theory}

Very recently, Aharony, Bergman, Jafferis and Maldacena (ABJM) have
constructed three-dimensional $U(N)\times U(N)$ and $SU(N)\times SU(N)$
Chern-Simons gauge theories with ${\cal N}=6$ superconformal symmetry.
From the brane constructions, they argued that the $U(N)\times U(N)$
theory at Chern-Simons level $k$ describes the low-energy limit 
of $N$ M2-branes probing a $C^4/Z_k$ singularity. It was 
conjectured that for $k=1$ and $2$, the ABJM theory respectively 
describes $N$ M2-branes in flat space and on a $R^8/Z_2$ orbifold,
and then may have ${\cal N}=8$ supersymmetry.
For $N=2$, this theory has additional symmetries and becomes identical
to the BLG theory. 
In this subsection, we will study the novel Higgs mechanism 
in the ABJM theory.

Following the convention in Ref.~\cite{0806.1519}, we can write
the explicit Lagrangian in ABJM theory as follows
\begin{eqnarray}
 \label{c}
  {\mathcal L} &=& 
    2 K \epsilon^{\mu\nu\lambda} {\rm Tr} \left(
        A^{\prime}_\mu \partial_\nu A^{\prime}_\lambda 
+ \frac {2i}{3} A^{\prime}_\mu A^{\prime}_\nu A^{\prime}_\lambda
        - \hat{A}_\mu \partial_\nu \hat{A}_\lambda 
- \frac {2i}{3} \hat{A}_\mu \hat{A}_\nu \hat{A}_\lambda \right)
     \nonumber \\
&&~~~~ - {\rm Tr} \left(({\mathcal D}_\mu Z)^\dagger {\mathcal D}^\mu Z
    +  ({\mathcal D}_\mu W)^\dagger {\mathcal D}^\mu W
    - i  \zeta^\dagger \gamma^\mu{\mathcal D}_\mu \zeta
    - i  \omega^\dagger \gamma^\mu{\mathcal D}_\mu \omega \right)
\nonumber \\
 &&~~~~- V_{\mathrm{ferm}} - V_{\mathrm{bos}}~,~\,
\end{eqnarray}
 where 
\begin{eqnarray}
K~=~{k\over {8 \pi} }~,~\,
\end{eqnarray}
\begin{eqnarray}
  Z^1  = X^1 + i X^5 ~,~~ ~Z^2  = X^2 + i X^6 ~,~~ 
  ~W_1  = X^{3\dagger} + i X^{7\dagger}  ~,~~ 
  ~W_2  = X^{4\dagger} + i X^{8\dagger}  ~,~\,
\end{eqnarray}
where $X^i$ belongs to the bifundamental representation  of $U(N)\times
U(N)$ or  $SU(N)\times SU(N)$,  
and here we do not present the potential $V_{\mathrm{ferm}}$ and $
V_{\mathrm{bos}}$ since they  are irrelevant in the following discussions.
For our convention, we choose
\begin{eqnarray}
{\rm Tr}(T^a T^b)~=~{1\over 2} \delta_{ab}~,~~~[T^a, ~T^b]~=~if_{abc} T^c~,~\,
\end{eqnarray}
where $T^{a, b, c}$ are the generators of the corresponding gauge group.

Similar to the novel Higgs mechanism in the BLG theory, we give the 
diagonal VEV to $X^8$ as follows
\be 
\langle X^8 \rangle  ~=~v_0 I_{N\times N}~=~v {\sqrt K} I_{N\times N}~,~\,
\ee 
where $I_{N\times N}$ is the $N$ by $N$ indentity matrix.
Also, we define
\begin{eqnarray}
A_\mu={1\over 2}(A^{\prime}_\mu+\hat{A}_\mu)~,~~~
B_\mu={1\over 2}(A^{\prime}_\mu-\hat{A}_\mu)~.~\,
\end{eqnarray}
So we have 
\begin{eqnarray}
A^{\prime}_\mu=A_\mu+B_\mu~,~~~\hat{A}_\mu=A_\mu-B_\mu~.~\,
\end{eqnarray}
From the kinetic term for $W_2$ and  the Chern-Simons terms, we obtain
the relevant Lagrangian for pure Yang-Mills fields
\be 
{\mathcal L}_{\rm YM} =   K \left(-2v^2B^a_\mu B^\mu_a+2 \epsilon^{\mu\nu\lambda} B^a_\mu
F_{a\nu\lambda}-\frac{2}{3} \epsilon^{\mu\nu\lambda} f_{abc} B^a_\mu B^b_\nu
B^c_\lambda \right)~,~
\label{ABJM-LAG}
\ee 
where  $F_{\mu\nu}=\partial_\mu A_\nu-\partial_\nu A_\mu+i[A_\mu,A_\nu]$.
Note that the BLG theory with $SO(4)$
gauge group is the same as the ABJM theory with $SU(2)\times SU(2)$
gauge group, so we can obtain the Lagrangian
in Eq. (\ref{a}) from that in
the above Eq. (\ref{ABJM-LAG}) by rescaling $f_{abc}$.

\section{ Effective Action for the Pure Gauge Fields }

Because $B_{\mu}^a$ is massive, we will calculate the effective action
for pure Yang-Mills fields by integrating it out. Due to the absence of 
the cubic term for $B_{\mu}^a$ in the (gauged) BF membrane theory, 
 we do not have the high-order corrections in
the effective action of gauge fields. Thus,
we will concentrate on the BLG theory and ABJM theory.
The relevant Lagrangians for pure gauge fields are the same
for the BLG theory and the ABJM theory with $SU(2)\times SU(2)$
gauge symmetry, and the ABJM theory is more general. Thus,
we will use the Lagrangian in
 Eq. (\ref{ABJM-LAG}) in the following discussions.

From the Lagrangian in Eq. (\ref{ABJM-LAG}), we get 
the equation of motion for $B^a_\mu$ 
\begin{eqnarray}
B^\mu_a~=~\frac{1}{2v^2} \epsilon^{\mu\nu\lambda} F_{a\nu\lambda}
-\frac{1}{2v^2} \epsilon^{\mu\nu\lambda} f_{abc} B^b_\nu B^c_\lambda~.~\,
\label{EOM-B}
\end{eqnarray}
We can solve the above equation by parametrizing the solution in
${1/{v^2}}$ expansion
\begin{equation}
B^\mu_a=\sum_n{1\over {v^{2n}}}(C_{2n})_a^{\mu}~.~\,
\end{equation}
Substituting it back into Eq. (\ref{EOM-B}), we obtain
\begin{equation} 
\label{aa}
\sum_n{1\over {v^{2n}}} (C_{2n})^\mu_a~=~{1\over
{2v^2}}\epsilon^{\mu\nu\lambda}F_{a \nu \lambda}-{1\over
{2v^2}}\sum_{n,m}{1\over v^{2n+2m}}{\epsilon^{\mu \nu \lambda}}
 f_{abc}(C_{2n})_\nu^b(C_{2m})_\lambda^c + ...~.~\,
\end{equation}

Because we only know for sure the high-order terms up to the order 
of $F^4$ in D2-brane effective 
action~\cite{Tseytlin:1997csa, Hashimoto:1997gm, Schwarz:2001ps}, 
we only need to calculate the 
solution to Eq. (\ref{EOM-B}) up to the order of $1/v^{10}$ or $(C_{10})^\mu_a$.
And the non-vanishing terms in the solution are
\begin{eqnarray}
(C_2)^\mu_a~=~{1\over2}{\epsilon^{\mu \nu \lambda}}F_{a \nu \lambda}~,~~~
(C_6)^\mu_a~=~-{1\over2}{\epsilon^{\mu \nu \lambda}}f_{abc}(C_2)_{\nu}^b(C_2)_\lambda^c~,~\,
\label{EQ-C-A}
\end{eqnarray}
\begin{eqnarray}
(C_{10})^\mu_a=-{\epsilon^{\mu\nu\lambda}}f_{abc}(C_2)_{\nu}^b(C_6)_\lambda^c~.~\,
\label{EQ-C-B}
\end{eqnarray}

Integrating $B^a_\mu$ out, we get the Lagrangian for pure Yang-Mills
fields
\begin{eqnarray}
{\mathcal L_{\rm YM}} &=& {\mathcal L^{(2)}_{\rm YM}} +   {\mathcal L^{(3)}_{\rm YM}} 
+  {\mathcal L^{(4)}_{\rm YM}}
+ ... ~,~\,
\end{eqnarray}
where
\begin{eqnarray}
{\mathcal L^{(2)}_{\rm YM}}
~=~\frac{2K}{v^2} (C_2)_{\mu}^a (C_2)^{\mu}_a~,~\,
\end{eqnarray}
\begin{eqnarray}
{\mathcal L^{(3)}_{\rm YM}} 
~=~ -\frac{2K}{3v^6}\epsilon^{\mu\nu\lambda}f_{abc}
(C_2)_{\mu}^a (C_2)_{\nu}^b (C_2)_{\lambda}^c~,~\,
\end{eqnarray}
\begin{eqnarray}
{\mathcal L^{(4)}_{\rm YM}}
~=~ -\frac{2K}{v^{10}} (C_6)_{\mu}^a (C_6)^{\mu}_a
-\frac{2K}{v^{10}}\epsilon^{\mu\nu\lambda}f_{abc} 
(C_2)_{\mu}^a (C_2)_{\nu}^b (C_6)_{\lambda}^c ~.~\,
\end{eqnarray}

Using Eqs. (\ref{EQ-C-A}) and (\ref{EQ-C-B}),
and the useful identities in the Appendix A, we obtain
\begin{eqnarray}
{\mathcal L^{(2)}_{\rm YM}}
 ~=~-\frac{2K}{v^2} {\rm Tr} \left(F^2 \right)  ~,~\,
\end{eqnarray}
\begin{eqnarray}
 {\mathcal L^{(3)}_{\rm YM}} 
~=~\frac{i4K}{3v^6}{\rm Tr}\left(F_{\alpha_1 \beta_1
}[F^{\beta_1\beta_3},F^{~~\alpha_1}_{\beta_3}] \right)
~,~\,
\end{eqnarray}
\begin{eqnarray}
{\mathcal L^{(4)}_{\rm YM}}
~=~ \frac{K}{2v^{10}}{\rm Tr} \left( [F^{\rho
\sigma},F^{\eta\delta}][F_{\eta\delta},F_{\rho \sigma}] \right)
 ~.~\,
\end{eqnarray}
Thus, ${\mathcal L^{(2)}_{\rm YM}}$
 is the kinetic term for the gauge fields
$A^\mu_a$ and is the leading order of the supersymmetric Yang-Mills 
 effective action. Moreover, the gauge coupling in the
BLG theory is 
\begin{eqnarray}
g_{YM}^2 ~=~ {{fv^2}\over 4}~=~ {{f^2v_0^2}\over 4} 
~\propto~ {{v_0^2}\over {k^2}}
 ~,~\,
\end{eqnarray}
and the  gauge coupling in the ABJM theory is
\begin{eqnarray}
g_{YM}^2 ~=~ {{v^2}\over {4K}}~=~ {{v_0^2}\over {4K^2}} 
~\propto~ {{v_0^2}\over {k^2}}
 ~.~\,
\end{eqnarray}
So for very large $v_0$ and $k$, we can still keep
the gauge coupling as a fixed constant.
For D2-branes, the gauge coupling is related to the string
coupling and the string length as follows
\begin{equation}
g_{YM}=({g_s\over \ell_s})^{1\over2}~.~\,
\end{equation}
And then for the fixed string coupling, we have
$g_{YM}^2\propto \alpha^{-1/2}$. 
Therefore,  ${1/v}$ is proportional to 
${\alpha}^{\prime {1/4}}$,  $ {\mathcal L^{(3)}_{\rm YM}} $ 
and ${\mathcal L^{(4)}_{\rm YM}}$
are proportional to $g_{YM}^{-2}\alpha^{\prime}$ and 
$g_{YM}^{-2}\alpha^{\prime 2}$, respectively. In short, they are 
at the correct orders according to the $\alpha'$ expansion.

Because ${\mathcal L^{(3)}_{\rm YM}} $ and ${\mathcal L^{(4)}_{\rm YM}}$
only have commutator terms, these high-order terms are covariant 
derivative terms and then do not contribute to the effective action for 
the D2-branes~\cite{Tseytlin:1997csa}. We conjecture that  
all the high-order terms obtained by this approach 
are the commutator terms. The point is
that the equation of motion for $B^a_\mu$ in Eq. (\ref{EOM-B})
can be rewritten as follows 
\begin{eqnarray}
B^{a\mu}~=~\frac{1}{2v^2} \epsilon^{\mu\nu\lambda} F^a_{\nu\lambda}
+\frac{i}{v^2} \epsilon^{\mu\nu\lambda} {\rm Tr} \left(
T^a [B_\nu, B_\lambda] \right)~.~\,
\label{EOM-CMT}
\end{eqnarray}
Because all the high-order terms originally come from the last term 
in the above equation which is a commutator term, all the high-order
terms should be the commutator terms and then the covariant terms.
Thus, moduloing the commutator terms or covariant derivative
terms, we only have  the kinetic term for the gauge fields
$A^\mu_a$ from the BLG and ABJM theories, which is the leading order 
in the D2-brane effective action. And then the effective action for
 pure Yang-Mills fields from the BLG and ABJM theories is the same 
as that from the (gauged) BF membrane theory after we integrate $B^\mu_a$
out. Therefore, how to obtain the non-trivial
$F^4$ terms in the D2-brane effective action from the 
BLG theory, the (gauged) BF membrane theory, and the ABJM theory
is still a big problem.

\section{D2-Branes from the Curved M2-Branes}

In spired by the derivation of a single D2-brane from a 
M2-brane~\cite{Bergshoeff:1996tu}, we would like to consider the
multiple curved M2-branes. To employ the 
trick in Ref.~\cite{Bergshoeff:1996tu}, we only need to introduce
 gravity. For simplicity, we do not consider the dilaton, the vector 
and scalar fields in the eleven-dimensional metric
 due to compactification, and RR fields, etc. And our ansatz for
 the Lagrangian of the curved M2-branes is
\be
\label{4a} 
{\mathcal L}_{\rm Curved}=-\beta_0 \sqrt{-{\rm det}(g)}
+\sqrt{-{\rm det}(g)}~{\mathcal L}_{\rm M2s}~,~\,
\ee
where $\beta_0$ is a positive constant like membrane tension, 
$g_{\mu \nu}$ is the induced metric on the world-volume of
 multiple M2-branes, and ${\mathcal L}_{\rm M2s}$ is formally given
in Eq. (\ref{LAG-BLG}) for the BLG theory
or in Eq. (\ref{c}) for the ABJM theory. 
In ${\mathcal L}_{\rm M2s}$, we need to
replace $\eta_{\mu \nu}$ and $\partial_\alpha$  
by $g_{\mu \nu}$  and $\nabla_\alpha$, respectively.
Also, we replace $\epsilon_{\mu \nu \lambda}$ by $\varepsilon_{\mu \nu
\lambda}=\sqrt{-g} \epsilon_{\mu \nu \lambda}$ which will be
covariant under coordinate transformation. This is a natural action
for the multiple M2-branes in the curved space-time
since it can come back to flat theory after we decouple the gravity.

Similar to the discussions in Ref.~\cite{Bergshoeff:1996tu}, we 
introduce an auxiliary filed $u$ and rewrite the above 
Lagrangian as follows
\be
\label{4b} 
{\mathcal L}_{\rm Curved}=\frac{\beta^2_0}{2u} {\rm det}~(g)-\frac{u}{2}
+\sqrt{-{\rm det}(g)}~{\mathcal L}_{M2s}~.~\, 
\ee
We can obtain the Lagrangian in Eq. (\ref{4a}) from
Eq. (\ref{4b}) by integrating out the auxiliary filed $u$.

To match the convention in~\cite{Tseytlin:1997csa},
we give the following VEV to the  scalar field $\phi$  
\be
\label{4c}
<\phi>=(\frac{8u}{\sqrt{-{\rm det}(g)}})^{1/2} ~{{K'}\over {\beta_0}}
~{1\over {2\pi \alpha'}}~  I_{N\times N}~,~\,
\ee 
where we can take
$\phi=X^{8(\phi)}$, $K'=1/f$, and $N=1$ in the BLG theory, 
take $\phi=X_+^{8}$, $K'=1$  and $N=1$ 
in the (gauged) BF membrane theory, and take
$\phi=X^{8}$ and $K'=K$ in the ABJM theory.
Thus, the relevant Lagrangian is 
\begin{eqnarray}
{\mathcal L}_{\rm Curved}&=&\frac{\beta^2_0}{2u} {\rm det}~
(g)-\frac{u}{2}+\sqrt{-{\rm det}(g)}\left(-2\langle \phi^2 \rangle  B^a_\mu B^\mu_a
+2K'\varepsilon^{\mu \nu \lambda} B_\mu^a F_{\nu \lambda}^a
\right.\nonumber\\&& \left.
-\frac{2}{3}K'\varepsilon^{\mu \nu \lambda}f_{abc} B_\mu^a
B_\nu^b B_\lambda ^c \right)~.~\, 
\end{eqnarray}

Using the results of the novel Higgs mechanism in the Section III
and neglecting the commutator terms for the $A_{\mu}^a$ field strength, 
we obtain
\be 
{\mathcal L}=\frac{\beta^2_0}{2u}
{\rm det}(g)\left(1+\frac{(2\pi \alpha')^2}{4} F^{a\mu\nu} F^a_{\mu \nu} \right)
-\frac{u}{2}~.~\,
\ee 

Moreover, we  
 use the following identity for the $3\times 3$ matrices that is proved in
the Appendix A
\be
\label{4d} 
{\rm Str} ~  {\rm det} ~(g+ 2\pi \alpha'{ F})
~=~{\rm det} (g)~\left(1+\frac{(2\pi \alpha')^2}{4} F^{a\mu\nu} 
F^a_{\mu \nu}\right)~,~\,
\ee
where ``Str'' is the symmetrized trace that acts on the gauge group indices,
and  ``det'' acts on the world-volume coordinate indices.
Integrating out  the auxiliary field $u$, we obtain
the Lagrangian for multiple D2-brane effective action
\begin{eqnarray}
{\mathcal L}=-\beta_0 {\sqrt {-{\rm Str} ~  {\rm det} ~(g+ 2 \pi \alpha' { F}) }}~.~\,
\label{D2-Action-I}
\end{eqnarray}

However, the well-known Lagrangian for the multiple D2-brane DBI action 
is~\cite{Tseytlin:1997csa}
\begin{eqnarray}
{\mathcal L}=-c_0{\rm Str}\left[\sqrt {- ~  {\rm det} ~
(g+ 2 \pi \alpha' { F}) }\right]~,~\,
\label{D2-Action-II}
\end{eqnarray}
where $c_0$ is a constant.
Because in general the Lagrangian in Eq. (\ref{D2-Action-I}) is not 
equivalent to that in Eq. (\ref{D2-Action-II}),
 we still can not get the correct $F^4$ terms for generic case.

Interestingly, for gauge symmetry $SU(2)\times SU(2)$ 
in the BLG theory, or the (gauged) BF membrane theory, or
 the ABJM theory, we indeed can get the correct $F^4$ terms.
Let us prove it in the following.
From the Lagrangian in Eq. (\ref{D2-Action-I}), we obtain
\begin{eqnarray}
{\mathcal L}=-\beta_0{\sqrt {-{\rm det}(g)\left(1+\frac{(2\pi \alpha')^2}{4} 
F^{a\mu\nu} F^a_{\mu \nu} \right)  }}~.~\,
\label{SO4-A}
\end{eqnarray}
Expandind the above Lagrangian, we have the relevant 
Lagrangian for pure Yang-Mills fields at the Minkowski space-time
limit
\begin{eqnarray}
{\mathcal L}=-{{\beta_0 (2\pi \alpha')^2} \over {4}} {\rm Tr} 
\Biggl[ F^{\mu \nu}  F_{\mu \nu}
 - {{(2\pi \alpha')^2}\over {4}}
(F^{\mu \nu}  F_{\mu \nu})^2 + ...\Biggr]~.~\,
\label{SO4-B}
\end{eqnarray}

From the known effective action for multiple D2-branes,
the relevant Lagrangian for pure Yang-Mills fields up to 
the $F^4$ terms is~\cite{Tseytlin:1997csa}
\begin{eqnarray}
{\mathcal L}_{DBI} & = & c_1 {\rm Tr} \Biggl\{  F^{\mu \nu}  F_{\mu \nu}
-{1\over 3} (2\pi \alpha')^2  \left(F^{\mu \nu} F_{\rho \nu}
F_{\mu \sigma}  F^{\rho \sigma} + \frac{1}{2} 
F^{\mu \nu} F_{\rho \nu} F^{\rho \sigma} F_{\mu \sigma} 
\right.\nonumber\\&& \left.
- \frac{1}{4}  F^{\mu \nu}  F_{\mu \nu}  F^{\rho \sigma}  F_{\rho \sigma}
-\frac{1}{8} F^{\mu \nu} F^{\rho \sigma} F_{\mu \nu}  F_{\rho \sigma}
\right)  + ... \Biggr\} ~,~\,
\label{SO4-C}
\end{eqnarray}
where $c_1=\pi^2 \alpha^{\prime 2} c_0$.
For gauge group $SU(2)$, we obtain
\begin{eqnarray}
{\mathcal L}_{DBI} & = &  c_1 {\rm Tr} \Biggl\{  F^{\mu \nu}  F_{\mu \nu}
- (2\pi \alpha')^2  \left(\frac{1}{8} (F^{\mu \nu} F_{\mu \nu})^2
 + \frac{1}{24} 
 F^{\mu \nu} F^{\rho \sigma} [F_{\mu \nu},  F_{\rho \sigma}]
\right)  + ... \Biggr\} ~.~\,
\label{SO4-D}
\end{eqnarray}
Therefore, neglecting the commutator terms and 
 rescaling the gauge fields,
we can show that the correct $F^4$ terms in the effective
D2-brane action in Eq. (\ref{SO4-B}) from the two M2-branes 
in the BLG and ABJM theories are equivalent to
these in the known DBI action in Eq. (\ref{SO4-D}).

In short, we can generate the correct $F^4$ terms in the 
effective D2-brane action from 
the  BLG theory and the ABJM theory with 
gauge group $SU(2)\times SU(2)$. However, we can not
get the  correct $F^4$ terms from  the ABJM theory
with $U(N)\times U(N)$ and $SU(N)\times SU(N)$ 
gauge symmetries for $N>2$. It seems to us
that the reasons are the following:  the BLG theory 
and the ABJM theory with gauge group $SU(2)\times SU(2)$ 
have three-dimensional ${\cal N}=8$ superconformal symmetry
while  the  ABJM theory with 
$U(N)\times U(N)$ and $SU(N)\times SU(N)$ 
gauge symmetries for $N>2$ might only have
 three-dimensional ${\cal N}=6$ superconformal 
symmetry~\cite{Bagger:2008se}.
However, for the (gauged) BF membrane theory, 
although  the constraint $\nabla_{\mu}X_+^8=0$ is still
satisfied,
it might be equivalent to 
three-dimensional ${\cal N}=8$ supersymmetric Yang-Mills theory.
In particular, for the (gauged) BF  membrane theory with 
$SU(2) \times SU(2)$ gauge symmetry, 
 we can generate the correct $F^4$ terms since
it is similar to the corresponding BLG and ABJM theories.

\section{Discussion and Conclusions}

Using the novel Higgs mechanism, we considered  the derivation
of the multiple D2-brane effective action for pure Yang-Mills
fields from the multiple 
M2-branes in the BLG theory and the ABJM theory. We showed
that the high-order $F^3$ and $F^4$ terms are commutator 
terms, and we argued that all the high-order terms are commutator 
terms as well. Thus, these high-order terms do not contribute
to the  multiple D2-brane effective action. 
In order to generate the non-trivial high-order $F^4$ terms and
inspired by the derivation of one D2-brane from one M2-brane,
we considered the curved M2-branes and introduce an auxiliary
field. In particular, the VEV of the scalar field in the novel 
Higgs mechanism depends on the auxiliary field. After we
integrate out the massive gauge fields and auxiliary field, 
we obtain the correct high-order 
$F^4$ terms in the D2-brane effective action from the BLG  theory
and the ABJM theory with $SU(2)\times SU(2)$ gauge group.
However, we still can not
obtain the correct $F^4$ terms in the generic ABJM theory
with gauge groups $U(N)\times U(N)$ and $SU(N)\times SU(N)$ 
for $N>2$. This might be related 
to the possible fact that the $SU(2)\times SU(2)$ gauge theory has 
three-dimensional ${\cal N}=8$ superconformal symmetry while the 
$U(N)\times U(N)$ and
$SU(N)\times SU(N)$ gauge theories for $N > 2$ might only have 
three-dimensional ${\cal N}=6$ superconformal symmetry. We also briefly
comment on the (gauged) BF membrane theory.

\begin{acknowledgments}

 This research was supported in part  by the
Cambridge-Mitchell Collaboration in Theoretical Cosmology (TL).

\end{acknowledgments}

\appendix

\section{Mathematical Identifies}

In this appendix we collect or prove the useful identities in this
paper:

(1) Two useful identities about $\epsilon$ in three-dimensional
Minkowski space-time
\bea
\epsilon^{\mu\nu\lambda}{\epsilon_{\lambda}}^{\rho\sigma}
=(-\eta^{\mu\rho}\eta^{\nu\sigma}+
\eta^{\mu\sigma}\eta^{\nu\rho})~,~\, 
\eea
\bea
\epsilon^{\mu\gamma\lambda}\epsilon^{\nu\rho\sigma}
=-\eta^{\mu\nu}(\eta^{\gamma\rho}\eta^{\lambda\sigma}
-\eta^{\gamma\sigma}\eta^{\lambda\rho})+\eta^{\mu\rho}
(\eta^{\gamma\nu}\eta^{\lambda\sigma}-
\eta^{\gamma\sigma}\eta^{\lambda\nu})-\eta^{\mu\sigma}
(\eta^{\gamma\nu}\eta^{\lambda\rho}-\eta^{\gamma\rho}
\eta^{\lambda\nu})~.~
\eea

(2) Let us prove the identity in Eq. (\ref{4d}) which is right for the
Abelian and non-Abelian cases
\bea {\rm Str} ~{\rm det} (g+ a { F})&=& 
{\rm det} (g_{\mu\nu})~ {\rm Str} ~{\rm det} (g^\nu_{~~\lambda}
+ a { F}^\nu_{~~\lambda})\nonumber \\
&=&({\rm det} ~g)~{\rm Str}
~\epsilon_{\alpha_1\alpha_2\alpha_3}(g^1_{~~\alpha_1}+ a { F}^{1}_{~~\alpha_1})
(g^2_{~~\alpha_2}+ a { F}^{2}_{~~\alpha_2})
(g^3_{~~\alpha_3}+ a { F}^{3}_{~~\alpha_3})
\nonumber \\
&=& ({\rm det} ~g)~{\rm Str} ~\Biggl[1+ a {F}^{\alpha}_{~~\alpha}+ a^2 
\left(\epsilon_{1\alpha_2\alpha_3}{ F}^2_{~~\alpha_2}
{F}^3_{~~\alpha_3}+ \epsilon_{\alpha_1 2\alpha_3} 
{F}^1_{~~\alpha_1} { F}^3_{~~\alpha_3}
\right.\nonumber\\&& \left.
~~+\epsilon_{\alpha_1\alpha_2 3}
{ F}^1_{~~\alpha_1} {F}^2_{~~\alpha_2} \right)
+ a^3
\epsilon_{\alpha_1\alpha_2\alpha_3}
{ F}^{1}_{~~\alpha_1} { F}^{2}_{~~\alpha_2} { F}^{3}_{~~\alpha_3}
\Biggr]\nonumber \\
&=& ({\rm det}~g)~(1+\frac{a^2}{4} F^{a\mu \nu} F^a_{\mu \nu} )~,~\,
\eea
where $a=2 \pi \alpha'$.

\renewcommand{\Large}{\large}

\end{document}